# A Growth Model with Unemployment

*Mina Mahmoudi*
Department of Economics
University of Nevada, Reno
1664 N Virginia St.
Reno, NV USA 89557
(775) 303-2535
mmahmoudi@unr.edu

*Mark Pingle*
Department of Economics
University of Nevada, Reno
1664 N Virginia St.
Reno, NV USA 89557
(775) 784-6634
pingle@unr.edu

**ABSTRACT**
A standard growth model is modified in a straightforward way to incorporate what Keynes (1936) suggests in the "essence" of his general theory. The theoretical essence is the idea that exogenous changes in investment cause changes in employment and unemployment. We implement this idea by assuming the path for capital growth rate is exogenous in the growth model. The result is a growth model that can explain both long term trends and fluctuations around the trend. The modified growth model was tested using the U.S. economic data from 1947 to 2014. The hypothesized inverse relationship between the capital growth and changes in unemployment was confirmed, and the structurally estimated model fits fluctuations in unemployment reasonably well.

**Keywords**
Growth theory, unemployment

## 1 INTRODUCTION

It is ironic that modern growth theory has primarily become a tool used to examine long term trends for full employment economies, for the first dynamic growth models were developed to examine whether full employment could be maintained as economies grew. Harrod (1939) identified his earlier work in "trade cycle theory" as what motivated him to develop his ground breaking model of growth. He sought to understand whether the income generating function of investment could be compatible with its capacity generating function. Similarly, Domar (1946, p. 139) emphasized the "dual character" of investment, noting, "because investment in the Keynesian system is merely an instrument for generating income, the system does not take into account the extremely essential, elementary and well-known fact that investment also increases productive capacity." Domar (1946, p. 138) sought "the rate of growth at which the economy must expand in order to remain in a continuous state of full employment." He found that "the failure of the economy to grow at the required rate creates unused capacity and unemployment" (Domar, 1946, p. 143).

Solow (1956, p. 65) summarized the work of Harrod and Domar in the following way: "The characteristic and powerful conclusion of the Harrod-Domar line of thought is that, even for the long run, the economic system is at best balanced on a knife-edge of equilibrium growth." Only by chance is the economy at full employment on a balanced growth path, and this is when the economy's savings rate happens to be a particular level relative to other parameters. Solow's (1956) primary "contribution to the theory of economic growth" was to show the Harrod-Domar knife-edge stems from an overly restrictive modeling assumption. Solow (1956) (and also Swan (1956)) showed full employment on a balanced growth path is possible under a wide variety of conditions if labor and capital are substitutable in production, rather than being restricted to a fixed ratio as assumed by Harrod and Domar.

In his Nobel address, Solow (1988) laments the focus of growth theory on full employment. He summarizes the history of its development by saying, "Growth theory was invented to provide a systematic way to talk about and to compare equilibrium paths for the economy," but then goes on to say, "in doing so, it failed to come to grips adequately with an equally important and interesting problem: the right was to deal with deviations from equilibrium growth" (Solow, 1988, p. 311). A primary message of his Nobel lecture was the "theory of equilibrium growth badly needed—and still needs—a theory of deviations from the equilibrium growth path" (Solow 1988, p309). This paper, in the most straightforward and direct manner, seeks to address this need.

## 2 A CANONICAL GROWTH THEORY MODEL

Standard growth theory assumes the economy's level of production depends on the economy's employment level, capital level, and level of technology. Let

(1) $Y = F(AL^d, K)$

be the production function for the model economy, where the employment level is that labor demand level $L^d$, the capital level is the capital demand level $K$, and the level of technology is $A$.

Technological growth is not explained by the model, but rather grows exogenously at the rate $a$ so the change in technology may be given by

(2) $A' = aA$.

Technological change is embodied in labor so $AL^d$ is the economy's effective labor level. Effective labor and capital are productive, meaning $F_{AL^d} > 0$ and $F_K > 0$, but are subject to diminishing returns, meaning $F_{AL^d, AL^d} < 0$ and $F_{KK} < 0$. Production exhibits constant returns to scale, so $\lambda Y = F(\lambda AL^d, \lambda K)$, where $\lambda$ is a scalar.

The economy is assumed to be competitive, with producers taking the nominal wage level $W$, the nominal capital rental rate $R$, and the price level $P$ as given. Assuming all producer optimization can characterized by a single representative firm with the production function (1), profit is maximized only if the labor demand level satisfies

(3) $\frac{W}{P} = F_{AL^d}(AL^d, K)$

and only if the capital demand level satisfies

(4) $\frac{R}{P} = F_K(AL^d, K)$.

That is, profit maximizing firms chose their labor levels so the real wage is equal to the marketing product of labor and the real capital rental rate is equal to the marginal product of capital.

The labor supply $L^s$ grows exogenously at the rate $n$ so

(5) $L^{s'} = nL^s$.

Here we introduce an unemployment measure $U$, given by the ratio

(6) $U = L^s/L^d$.

In the standard model, the nominal wage $W$ is determined by assuming it adjusts to equate labor supply and labor demand:

(7) $L^d = L^s$.

Because labor market clearing is assumed, the standard growth model does not distinguish labor demand from labor supply nor is an unemployment measure included. The "involuntary unemployment" described by Keynes as being equivalent to a labor surplus is ruled out by assumption.

Capital accumulates in the economy as savings flows into investment. Households are assumed to save the constant fraction $s$ of the real income generated from the sale of the output produced, so

(8) $S = sY$.

This supply of private saving finances private sector investment $I$, the public budget deficit $G - T$, and net exports $X$, so

(9) $S = I + [G - T] + X$.

The rental rate $R$ on capital is assumed to adjust, to eliminate a surplus or shortage of saving, so condition (9) holds. That is, like the labor market, the capital market is assumed to clear and remain in equilibrium over time. Capital depreciates at the rate $\delta$, so when investment and depreciating capital are each considered, the change in the capital level is given by

(10) $K' = I - \delta K$.

As presented in paper, data on the government budget deficit and net exports suggest that their levels remain roughly proportionate to the level of production. Here this will be modeled by assuming the level of taxes $T$ is the fraction $t$ of real income, the level of government purchases $G$ is the fraction $g$ of the real income, and the level of net exports $X$ is the fraction $x$ of income, so

(11) $T = tY$,
and
(12) $G = gY$,
and

(13) $X = xY$.

The implications of this model can be derived by reducing it to a single dynamic equation that describes the path of a single core state variable for the reduced form model. By defining $k^d = K/[AL^d]$, the basic dynamic equation for this standard growth theory model is

(14) $k^{d'} = (s + t - g - x)f(k^d) - (a + n + \delta)k^d$,

which describes the path of the core state variable $k^d$. In the appendix, it is also shown equations (1)-(13) also imply the following three auxiliary equations for the reduced form model:

(15) $y = f(k^d)$,
(16) $r = f'(k^d)$,
(17) $w = f(k^d) - f'(k^d)k^d$,

where the real wage is given by $w = \frac{W}{P}$, the real rate of return on capital is given by $r = \frac{R}{P}$, and the output level per effective unit of labor is given by $y = Y/[AL^d]$. With the initial value of $k^d$ given, equations (14)-(17) determine the paths for the variables $k^{d'}$, $y$, $r$, and $w$.

In summary, standard growth theory provides an explanation of long term economic trends. What the standard growth theory model cannot well explain is short term economic fluctuations, and it cannot at all explain involuntary unemployment. Deviations from the long term growth trend spawned by the constant rate of technical change can be explained by variations in the other exogenous variables (the capital depreciation rate, population growth rate, tax rate, rate of government spending, rate of net export spending). However, because labor demand is assumed to be equal to the exogenously growing labor supply, the standard growth theory model cannot explain fluctuations in unemployment. In the next section, we show that by relaxing the assumption of labor market clearing, a modified growth model can explain economic fluctuations and fluctuations in involuntary unemployment.

## 3 ALTERING THE GROWTH THEORY MODEL TO INCORPORATE THE ESSENCE OF KEYNES' GENERAL THEORY

When he converted the words of Keynes' (1936) "general theory" into a mathematical model, Hicks (1937) created what has become known as the IS-LM model. The IS-LM model captures the Keynesian notion that the economy, in the short term, is driven primarily by fluctuations in aggregate demand. In particular, fluctuations in the level of employment and involuntary unemployment depend upon fluctuations in aggregate demand.

Asset market activity and the LM curve are not necessary to develop what Keynes (1936, Chapter 3) called the "essence" of his general theory. That is, the essence of Keynes' general theory reduces to an IS equation and a production function which are used to show that fluctuations in the economy's employment level may arise from exogenous changes in investment demand.

The driving force of this Keynesian version of the growth model is the assumption that the path for capital is exogenously determined, and this driving force can be recognized in a intensive form for the model by introducing the variable $b = K'/K$, the growth rate of capital. If $K'$ is exogenously determined and $K$ predetermined for the levels model above, then the variable $b$ is exogenously determined for the intensive form model that will now be derived. Eliminating the variables in equation (9) using equations (8), (10), (11), (12), (13), and replacing the variable $Y$ using equation (1) yields $K' + \delta K = (s + t - g - x)F(AL^d, K)$. Dividing by $AL^d$ and remembering the definition $k^d = K/AL^d$, this latter equation becomes $[K'/K]k^d + \delta k^d = (s + t - g - x)[F(AL^d, K)]/AL^d$. The constant returns to scale assumption and the definition $F(1, k^d) = f(k^d)$ implies $F(AL^d, K)/[AL^d] = f(k^d)$, so using the definition $b = K'/K$, we have

(18)    $[b + \delta]k^d = [s + t - g - x]f(k^d)$.

Equation (18) provides a relationship between the intensive capital demand level $k^d$, the endogenous variable determined by the equation, and six other variables. Noticeably missing is the technology growth rate variable $a$, an indication that the rate of technical change does not impact the employment level through the variable $k^d$.

The change in the unemployment level depends not only on the change in intensive capital demand but also the change in the intensive capital supply. Since $U = L^s/L^d$, it also follows that $U = k^d/k^s$. Differentiating this condition, we obtain $U'/U = k^{d'}/k^d - k^{s'}/k^s$, which is same as

(19)    $U'/U = k^{d'}/k^d - [b - [a + n]]$.

When the capital growth rate is not constant, the growth rate of the unemployment level fluctuates. Differentiating condition (18), we obtain the following relationship between $b'$ and $k^{d'}$:

(20)    $k^{d'}/k^d = b'/[(s + t - g - x)f'(k^d) - (b + \delta)]$.

Using condition (20) to eliminate $k^{d'}/k^d$ from condition (19), we obtain

(21)    $\frac{U'}{U} = a + n - b + \frac{b'}{[s+t-g-x]f'(k^d)-[b+\delta]}$.

For the empirical work below, it is useful to restate condition (21). Using equation (18), the value $b + \delta$ is equal to $[s + t - g - x]\frac{f(k^d)}{k^d}$. Using this latter condition to replace $b + \delta$ in equation (21), we obtain $\frac{U'}{U} = a + n - b + \frac{b'}{\left[[s+t-g-x]\left[f'(k^d) - \frac{f(k^d)}{k^d}\right]\right]}$. The definition of the real wage in (17) then implies $\frac{U'}{U} = a + n - b - \frac{b'}{[s+t-g-x][w/k^d]}$. This is the same as

(22)    $\frac{U'}{U} = a + n - b - \frac{b'}{[s+t-g-x][WAL^d/PY][Y/K]}$.

In (22), $WAL^d/PY$ is the share output paid to labor, and $Y/K$ is the output to capital ratio.

## 4 FITTING THE MODEL TO THE DATA

In this section, we fit our suggested model to data for the U.S. from 1947 to 2104, and show the model has reasonable explanatory power. For the model presented in section 3, equation (22) relates the changes in the unemployment rate to the changes in the capital growth rate. The growth rate of technology $a$ is assumed constant. However, it may not be constant, (and it turns out that a changing rate best fits the data). To allow for a changing rate of technical change, assume the level of technology is given by $A = e^{a_0+a_1t+a_2t^2+a_3t^3}$ so the rate of technical change is $a = a_1 + 2a_2t + 3a_3t^2$. Replacing the variable $a$ in equation (22) and rearranging terms, we obtain

(23)    $\frac{U'}{U} - n + b =$
$a_1 + 2a_2t + 3a_3t^2 + \left[\frac{1}{[s+t-g-x][WAL^d/PY][Y/K]}\right][b']$.

Combining data for the variables on the left side of (23), we obtain a variable we can regress on the time variables and $b'$ to obtain an estimate for $\frac{1}{[s+t-g-x][WAL^d/PY][Y/K]}$. We expect this estimate to be negative. Intuitively, a higher rate of capital growth rate increases the demand for output, which increases the demand for labor, which curbs unemployment. Regressing $\frac{U'}{U} - n + b$ on $t$, $t^2$, and $b'$, we obtain

(24)    $\frac{U'}{U} - n + b = 0.023 + 0.00018t - 0.000008t^2 - 0.597b'$

where all coefficients are significant and $R^2 = 0.59$.

Using (24), we can construct the predicted paths for both the rate of change in unemployment $\frac{U'}{U}$ and the rate of technical change $a$. The significant estimate on the $t^2$ variable indicates the rate of technical change is not constant over time. Rather, our technical change indicates the rate of technical change follows the path $a = 0.023 + 0.00018t - 0.000008t^2$. The model $\frac{U'}{U}$ path is plotted in Figure 1, along with the actual path, and the model rate of technical change $a$ is plotted in Figure 2.

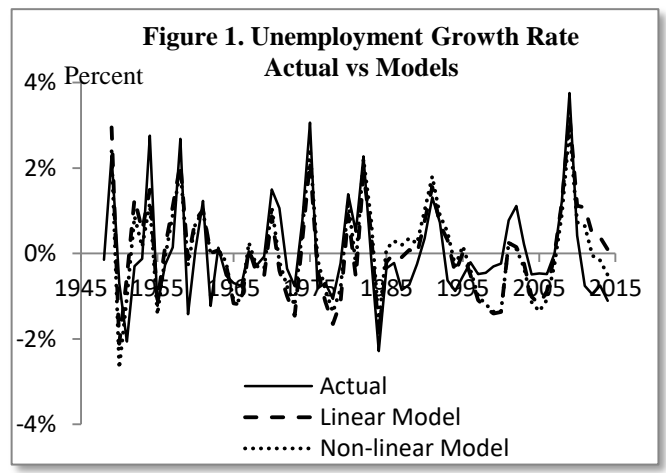

Figure 1. Unemployment Growth Rate Actual vs Models

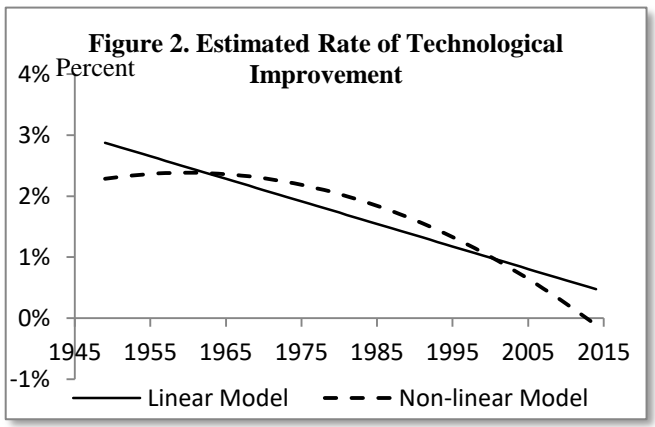

Figure 2. Estimated Rate of Technological Improvement

The path for technical change shown in Figure 13 shows an estimated negative rate of technical change for the present. While a declining rate of technical change is conceivable, a negative rate of change for the present is not realistic. To move to a more realistic model, we drop the $t^2$ variable and estimate a linear path for the rate of technical change. Regressing the dependent variable on b' and t and we obtain

(25) $\frac{U'}{U} - n + b = 0.029 - 0.00037t - 0.582b'$,

where all coefficients are significant and $R^2 = 0.54$.
In this case, the rate of technical change follows the path $a = 0.029 - 0.00037t$, and that path is plotted in Figure 2. The path for the unemployment growth rate $\frac{U'}{U}$ for this model is presented in Figure 1.

The model (25) indicates the rate of technical change as slowed over time, from 2.9 percent in 1949 to 0.5 percent in 2014. This path, along with the decreasing growth rate for capital, explains why the growth rate of output for the U.S. economy has decreased over time. That is, this model can explain the long term growth rate of the U.S. economy, like the standard growth theory model. However, as shown in Figure 1, this model effectively explains variations in unemployment.

## 5 CONCLUSION

We have shown standard growth theory can be adjusted in a natural way to incorporate the essence of Keynes' general theory, which hypothesizes that fluctuations in employment and unemployment are caused by exogenous changes in investment. Compared to the standard model, our modified growth model is unique in that it specifies the path for capital as being exogenously determined.\It is also relatively unique in that it relaxes the assumption of full employment and introduces a stick wage. In summary, the reader can think of our model as the IS-LM model of growth theory, where the money market has not been included.

In the empirical section, we showed our model explains unemployment movements relatively well. A byproduct of our structural estimation process was the finding that the rate of technological improvement for the economy is decreasing. Our primary finding is that capital growth and an increase in the capital growth rate each have a significant and positive impact on the change in the unemployment level. We find that the marginal impact of $b'$ (the change in the capital growth rate) on the unemployment growth rate $U'/U$ is -0.58, meaning an increase in the capital growth rate by one percent reduces the growth rate of unemployment by 0.58 percent.

Here we have restricted ourselves to examining how fluctuations in the capital growth rate impact fluctuations in unemployment. This cause and effect relationship was what Keynes called the essence of his general theory. As show in section 3, our model can be used to examine additional fluctuations, for example fluctuations in savings, tax, government spending, depreciation, and net export rates. While our work here has made clear that fluctuations in the capital growth rate explain much, further empirical work could examine how changes in other factors influence unemployment.